\documentstyle[here,epsfig]{mn2e}
\def\simlt{\mathrel{\rlap{\lower 3pt\hbox{$\sim$}}\raise 2.0pt\hbox{$<$}}}
\def\simgt{\mathrel{\rlap{\lower 3pt\hbox{$\sim$}} \raise 2.0pt\hbox{$>$}}}

\def\gtsima{$\; \buildrel > \over \sim \;$}
\def\ltsima{$\; \buildrel < \over \sim \;$}
\def\gtrsim{\lower.5ex\hbox{\gtsima}}
\def\lesssim{\lower.5ex\hbox{\ltsima}}
\def\url#1{{\ttfamily\def\/{/\discretionary{}{}{}}#1}}
\begin{document}

\newcommand{\q}{\begin{equation}}
\newcommand{\qa}{\begin{eqnarray}}
\newcommand{\qs}{\begin{eqnarray*}}
\newcommand{\nq}{\end{equation}}
\newcommand{\nqa}{\end{eqnarray}}
\newcommand{\nqs}{\end{eqnarray*}}
\newcommand{\ud}{\mathrm{d}}

\title[Dark matter decays and annihilations] 
{Impact of dark matter decays and annihilations on reionization}
\author[M. Mapelli, A. Ferrara and E. Pierpaoli]
{M. Mapelli$^{1}$,  A. Ferrara$^{1}$ and E. Pierpaoli$^{2}$\\
$^{1}$SISSA, International School for Advanced Studies, Via Beirut 4, I-34100, Trieste, Italy; {\tt mapelli@sissa.it}\\
$^{2}$Theoretical Astrophysics, California Institute of Technology, Pasadena, CA 91125, USA\\}

\maketitle \vspace {7cm }

\begin{abstract}
One of the possible methods to distinguish among various dark matter candidates is to study the effects of dark matter decays. We consider four different dark matter candidates (light dark matter, gravitinos, neutralinos and sterile neutrinos), for each of them deriving the decaying/annihilation rate, the influence on  reionization, matter temperature  and CMB spectra. We find that light dark matter particles (1-10 MeV) and sterile neutrinos (2-8 keV) can be sources of partial early reionization ($z\lesssim{}100$). However, their integrated contribution to Thomson optical depth is small ($\lesssim{}0.01$) with respect to the three year WMAP results ($\tau{}_e=0.09\pm{}0.03$). Finally, they can significantly affect the behavior of matter temperature. On the contrary, effects of heavy dark matter candidates (gravitinos and neutralinos) on reionization and heating are minimal. All the considered dark matter particles have completely negligible effects on the CMB spectra.

\end{abstract}

\section{Introduction}
The nature of dark matter (DM) is one of the crucial open questions in cosmology. In the so called cold dark matter (CDM) theory DM particles are defined as 'cold' particles, because of their negligible free-streaming length (i.e. the length  below which DM fluctuations are suppressed). The most famous alternative model to CDM is called warm dark matter (WDM), where DM particles are defined as 'warm' because of their longer free-streaming length. In WDM scenarios the velocity dispersion of the particles is sufficient to smear out  the fluctuations up to galactic scales, depending on the mass of the particles (Padmanabhan 1995). This means that WDM models can alleviate the so called substructure crisis, which represents one of the most serious problems of CDM theories (Bode, Ostriker \& Turok 2001; Ostriker \& Steinhardt 2003).
At present, there is no definitive evidence which allows us to exclude one of the two scenarios and even the properties (mass, lifetime, etc) of cold and warm dark matter particles are substantially unknown. 

From an observational point of view, one of the most direct ways to detect DM particles and, maybe, distinguish between CDM and WDM is represented by particle decays (Chen \& Kamionkowski 2004; Pierpaoli 2004). A small fraction of DM particles is expected to decay, the lifetime of this process generally depending on their density and mass. Many of the possible decay channels (Dolgov 2002) involve the emission of photons at wavelengths depending on the particle mass. So, in principle, it is possible to distinguish among various DM models, depending on the characteristics of emitted photons. It has also been pointed out (Sciama 1982;  Hansen \& Haiman 2004; Chen \& Kamionkowski 2004; Kasuya, Kawasaki \& Sugiyama 2004; Kasuya \& Kawasaki 2004; Padmanabhan \& Finkbeiner 2005; Mapelli \& Ferrara 2005; Zhang et al. 2006) that photons due to particle decays or annihilations can be sources of partial early reionization. In addition, the heating induced by particle decays can leave an imprint in the 21 cm background, which could be detected by new generation of radio telescopes (LOFAR, PAST, SKA, etc.).

At the moment, constraints on the radiation emitted by particle decays are loose. The SPI spectrometer aboard ESA's INTEGRAL satellite recently detected an excess in the 511 keV line emission, due to positron-electron annihilation, from the galactic bulge (Kn\"odlseder et al. 2005). The only two viable explanations are that this excess is due to positrons produced by thermonuclear Type Ia supernovae (Dermer \& Murphy 2001) or by decaying/annihilating dark matter (Hooper \& Wang 2004; Boehm et al. 2004). This last hypothesis, although exotic, has triggered many theoretical studies (Ascasibar et al. 2006; Kawasaki \& Yanagida 2005; Kasuya \& Takahashi 2005; Cass\'e \& Fayet 2005; Kasuya \& Kawasaki 2006), which are aimed to put constraints on various DM properties by using SPI/INTEGRAL observations.
 
In this paper, we consider some of the most popular cold and warm DM particles, calculating their approximate decaying rate (Section 2), their influence on the ionization fraction, on the Thomson optical depth, on the behavior of matter temperature (Sections 3 and 4) and on the cosmic microwave background (CMB) spectra (Section 5).\\
In all the cases, we make the assumption that the DM is composed of one single species of particles. We consider only 'standard' DM candidates, neglecting more exotic scenarios (such as Q-balls, light scalar bosons, etc). In particular, we study three different candidates for the case of CDM: (i) the axino, as representative of light dark matter (LDM; 1-100 MeV; Hooper \& Wang 2004), (ii) the gravitino and the (iii) neutralino, as heavy dark matter candidates ($\gtrsim{}100$ MeV). For the WDM we consider only the sterile neutrino, reducing our analysis to its radiative decay channel. In fact, we do not pretend to present a complete overview of DM candidates. Instead, we would like to give a basic description of the effects of DM decays, taking as an example some of the standard DM candidates.
 Our aim is to point out the differences among the considered DM particles, with particular care for cosmic reionization and heating.

\section{Method}
For each considered particle model we derived the energy injection rate per hydrogen nucleus, $\epsilon{}_{DM}$, as follows. 

In comoving coordinates the photon emission rate due to particle decay can be generally written as:
\q\label{eq:eq1}
\frac{{\ud}n}{{\ud}t}=\frac{n_0}{\tau{}}\,{}e^{[t(0)-t(z)]/\tau{}},
\nq
where $n_0$ and $\tau$ are the current density  and the lifetime of particles, respectively, and $t(z)\sim{} \frac{2}{3}H_0^{-1}\Omega{}_{0M}^{-1/2}(1+z)^{-3/2}$ (neglecting $\Omega_\Lambda{}$) is the time elapsed from the Big Bang to redshift $z$. 

Instead, in the case of annihilations, the photon emission rate in comoving coordinates is
\q\label{eq:eq1bis}
\frac{{\ud}n}{{\ud}t}=n_0^{2}\,{}(1+z)^3\sigma{}\,{}v\,{}C,
\nq
where $\sigma{}\,{}v$ is the annihilation cross-section (see Section 3.3 for details) and $C$ is the clumping factor.  To quantify $C$ is beyond the scope of this paper. However, especially at very high redshift ($z\gg{}10$), where annihilations play their most important role, we can roughly approximate $C\sim{}1$.

Then, in both the cases $\epsilon{}_{DM}$ is simply:
\q\label{eq:eq2}
\epsilon{}_{DM}=\frac{{\ud}n}{{\ud}t}\,{}\frac{E_\gamma{}}{n_b},
\nq
where $E_\gamma{}\sim{}m_{DM}/2$ is the energy of the emitted photon ($m_{DM}$ being the mass energy of the DM particle), and $n_b$ the current density of baryons (we take $n_b=2.7\times{}10^{-7}$ cm$^{-3}$, Spergel et al. 2003).

The energy $\epsilon{}_{DM}$ partially goes into ionizations of hydrogen and helium atoms and partially into heating.
We adopt the rough approximation by Chen \& Kamionkowski (2004), for which a fraction $(1-x)/3$ of  $\epsilon{}_{DM}$ contributes to ionizations and a fraction  $(1+2x)/3$ goes into heating ($x$ being the ionized fraction). This approximation is correct if the DM decays/annihilations produce (directly or via secondary interactions) photons
whose energy is mostly absorbed by the intergalactic medium in a Hubble time. As fig. 2 of Chen \& Kamionkowski (2004) shows, photons emitted by sterile neutrinos ($\sim{}1-4$ keV) fully satisfy this requirement. Particle decays/annihilations which produce electrons of energy lower than $\sim{}$1 GeV (Chen \& Kamionkowski 2004) equally satisfy this requirement, because these electrons can lead (by collisional ionization, excitation or inverse-Compton scattering) to photons which are immediately absorbed. Then, also for LDM and gravitinos we can adopt the approximation of Chen \& Kamionkowski (2004). For neutralinos (whose mass is higher than 30 GeV) this approximation is far too optimistic; however we can consider it as an upper limit.
In practice, in the optically thin case most of the decay/annihilation
products can propagate to redshift 0 and can be constrained by observing the
X-ray/gamma-ray background (Chen \& Kamionkowski 2004).

We made our calculations using an upgraded version of the public code RECFAST (Seager, Sasselov \& Scott 1999, 2000). In particular, we modified the evolution equations as follows (Padmanabhan \& Finkbeiner 2005).
\qa\label{eq:eq3}
-\delta{}\left(\frac{{\ud}x_H}{{\ud}z}\right)=\frac{\epsilon{}_{DM}}{E_{th,\,{}H}}\frac{1-x_H}{3\,{}(1+f_{He})}\,{}\mathcal{E} \\\label{eq:eq4}
-\delta{}\left(\frac{{\ud}x_{He}}{{\ud}z}\right)=\frac{\epsilon{}_{DM}}{E_{th,\,{}He}}\frac{1-x_{He}}{3\,{}(1+f_{He})}\,{}\mathcal{E}\\\label{eq:eq5}
-\delta{}\left(\frac{{\ud}T_{M}}{{\ud}z}\right)=\frac{2\,{}\epsilon{}_{DM}}{3\,{}k_B}\frac{1+2\,{}x_{H}+f_{He}\,{}(1+2\,{}x_{He})}{3\,{}(1+f_{He})}\,{}\mathcal{E},
\nqa
where $x_H$ ($x_{He}$) is the ionized fraction of hydrogen (helium) atoms, $E_{th,\,{}H}=13.6$ eV ($E_{th,\,{}He}=24.6$ eV) is the ionization energy of hydrogen (helium) atoms, $f_{He}$  is the helium-to-hydrogen ratio by number, $T_M$ is the matter temperature, $k_B$ the Boltzmann constant and  ${\mathcal E}\equiv{}\left[H(z)(1+z)\right]^{-1}$.

\section{Cold dark matter}
First we  consider CDM particles. Heavy CDM particles ($\geq{}$ 100 MeV) are not considered a viable source for the 511 keV emission in the galactic center. In fact, in the case of neutralinos (with mass higher than 30 GeV), the request of a sizable R-parity violation, needed to allow considerable neutralino decays, would determine a too short lifetime and the neutralino would cease to be a good DM candidate (Hooper \& Wang 2004). On the other hand, gravitino decays are possible; but the gravitino lifetime is far too long to match the 511 keV emission from the galactic center  (Hooper \& Wang 2004). Then, previous studies (Bohem et al. 2004; Hooper \& Wang 2004; Ascasibar et al. 2006) proposed light cold DM candidates (1-100 MeV) to be sources of the  511 keV emission from the galactic center.
 In the following, we will consider firstly LDM particles (axinos) and secondly heavy DM particles (gravitinos and neutralinos).

\subsection{Light dark matter}
LDM candidates (1-100 MeV) can produce positrons both via decay (axinos, Hooper \& Wang 2004) and via annihilation (Bohem et al. 2004). They can easily explain the 511 keV emission from the galactic center and satisfy the DM relic density ($\Omega_{DM}\sim{}0.23$, Spergel et al. 2003). The upper limit of their mass, if they are the source of the 511 keV line, is probably much less than 100 MeV, due to constraints on the bremsstrahlung emission (Beacom, Bell \& Bertone 2005; Cass\'e \& Fayet 2005; Beacom \& Y\"uksel 2006).

Hooper \& Wang (2004) derived in a very simple way the lifetime of decaying LDM particles (i.e. axinos) necessary to produce the observed 511 keV emission:
\q\label{eq:eq6}
\tau{}\sim{}4\times10^{26}\textrm{ s}\,{}\left(\frac{m_{LDM}}{\textrm{MeV}}\right)^{-1},
\nq
where $m_{DM}$ is the mass of a LDM particle. Under our assumptions, the current density of LDM particles can easily been derived as:
\q\label{eq:eq6bis}
n_0=\Omega_{DM}\,{}\frac{\rho{}_c}{m_{LDM}},
\nq
where $\rho{}_c$ is the critical density of the Universe.
Substituting  equation (\ref{eq:eq6}) and (\ref{eq:eq6bis}) into equation (\ref{eq:eq1}) and implementing it in RECFAST (through equations \ref{eq:eq3}-\ref{eq:eq5}), we derive the influence of LDM particles on ionization and heating, shown in Fig. 1. The contribution of LDM starts to be important at redshift $z\leq{}100$. The current value of $x_e$ should be of the order of 0.1 (for masses $m_{LDM}\gtrsim{}5$ MeV). Then LDM does not produce a complete reionization; but can be an important source of early partial ionization. Also the matter temperature $T_M$ starts to differ from the case without LDM at $z\sim{}100$. At $z\sim{}20$  $T_M$ is $\sim{}200$ K, a factor $\sim{}$20 higher than in the unperturbed case. At lower redshift  $T_M$ becomes of the order of 10$^4$ K. These high temperatures enhance the role of collisions (both between electrons and hydrogen atoms and between two hydrogen atoms). However, the collision time-scale (Palla, Salpeter \& Stahler 1983) remains always two or more orders of magnitude longer than the Hubble time, allowing us to neglect collisions in our calculations.

\begin{figure}
\center{{
\epsfig{figure=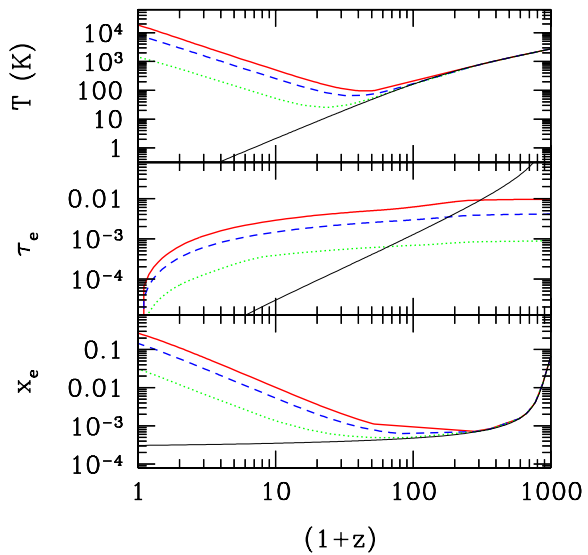,height=8cm}
}}
\caption{\label{fig:fig1}
Ionized fraction (bottom panel), Thomson optical depth (central panel) and matter temperature (upper panel) as a function of redshift due to decaying LDM of masses 1 (thick dotted line), 5 (dashed) and 10 MeV (solid). The thin solid line represents, from bottom to top, the relic fraction of free electrons, their contribution to Thomson optical depth and the matter temperature without particle decays.  
} 
\end{figure}
\subsection{Gravitinos}
The most probable gravitino masses are $m_{3/2}<1\,{}$keV and $m_{3/2}>1\,{}$TeV (Nowakowski \& Rindani 1995). In the first case, the gravitinos are the lightest supersymmetric particle (LSP) and then are stable: they are good WDM candidates; but their decay rate is negligible. In the second one, gravitinos are so unstable that they decay in the early universe, and, as a consequence, they are not viable DM candidates.
Some models assume that gravitinos have mass $m_{3/2}\sim{}10-100$ MeV, are the LSP and can violate the R-parity (Hooper \& Wang 2004). In this case gravitinos are good DM candidates and have non-negligible decay rate. For this case, we calculated the gravitino contribution to reionization and  heating, assuming lifetime:
\q\label{eq:eq7}
\tau{}\sim{}10^{31}\textrm{ s}\,{}\left(\frac{m_{\bar{l}}}{\textrm{100 GeV}}\right)^{4}\,{}\left(\frac{0.1\textrm{ GeV}}{m_{3/2}}\right)^7\,{}\left(\frac{0.1}{\lambda{}}\right)^2,
\nq 
where $m_{\bar{l}}$ is the slepton mass and $\lambda{}$ is the R-parity violating leptonic trilinear coupling. The current density of gravitinos can be derived as indicated for LDM. We find that, because of such a long lifetime, the contribution of gravitinos to heating and reionization is negligible (Fig. 2). For the same reason, Hooper \& Wang (2004) show that gravitinos are unable to produce the 511 keV excess from the galactic center.
\begin{figure}
\center{{
\epsfig{figure=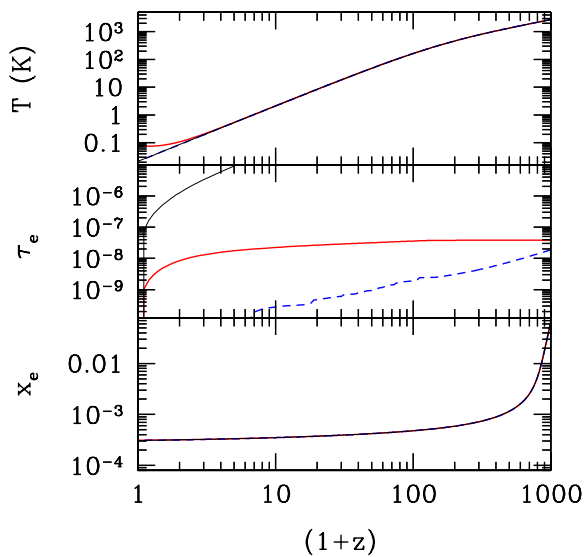,height=8cm}
}}
\caption{\label{fig:fig2} 
Ionized fraction (bottom panel), Thomson optical depth (central panel) and matter temperature (upper panel) as a function of redshift due to decaying gravitinos of masses 10 (dashed) and 100 MeV (solid). The thin solid line is the same as in Fig. 1.
} 
\end{figure}
\subsection{Heavy dark matter: neutralinos}
Here we will consider as 'heavy' dark matter the neutralinos. It is a merely indicative classification, given the uncertainties on the various models. The discussion of the details of different supersymmetric models is beyond the purpose of this paper. Neutralinos are thought to be very massive ($m_\chi{}>30$ GeV). So, if they could decay (violating the R-parity), their lifetime should be very short, and they could not be a viable DM candidate. Then, the neutralino, if exists, must be perfectly stable, and we will not treat neutralino decay. However, neutralinos can annihilate. The annihilation cross-section is generally fit by (Bertone, Hooper \& Silk 2005):
\q\label{eq:eq8}
\sigma{}\,{}v=a+b\,{}v^2+{\mathcal O}(v^4),
\nq
where $a$ and $b$ are constant, whose values are constrained by the DM relic density condition, and $v$ is the neutralino velocity, which depends on the DM temperature and thus on the redshift. In the present epoch neutralinos are non-relativistic, then the current annihilation cross-section can be written as $\sigma{}\,{}v\sim{}a$. However, the cross-section at the freeze-out time should depend on $v$ and be higher than the current value. As a rough approximation, Padmanabhan \& Finkbeiner (2005) consider a thermally averaged, redshift independent cross-section, $\langle{}\sigma{}\,{}v\rangle{}=2\times{}10^{-26}$ cm$^{3}$ s$^{-1}$. For comparison, we made the same assumption. Then, the annihilation rate becomes:
\qa\label{eq:eq9}
\frac{{\ud} n}{{\ud} t}=2.88\times{}10^{-42}\,{}(1+z)^3\,{}\textrm{cm}^{-3}\textrm{s}^{-1}\left(\frac{n_0}{1.2\times{}10^{-8}\textrm{cm}^{-3}}\right)^2\nonumber{}\\
\times{}\left(\frac{\langle{}\sigma{}\,{}v\rangle{}}{2\times{}10^{-26}\textrm{ cm}^3 \textrm{ s}^{-1}}\right),
\nqa
where $n_0=\Omega_{DM}\,{}\rho{}_c/m_\chi{}$. In our calculations we assume $m_\chi{}$=100 GeV. We have implemented this equation into RECFAST. The contribution of neutralino annihilations both to ionizations and heating is negligible (Fig. 3; dashed line) for  $\langle{}\sigma{}\,{}v\rangle{}=2\times{}10^{-26}$ cm$^{3}$ s$^{-1}$ (in agreement with Padmanabhan \& Finkbeiner 2005). As an upper limit, we considered also the case where $\langle{}\sigma{}\,{}v\rangle{}=10^{-24}$ cm$^{3}$ s$^{-1}$, which is the highest value to be consistent with the first year WMAP data (see Colafrancesco, Profumo \& Ullio 2005). Also in this case, the contribution to heating is negligible and the ionization fraction remains of the order of $10^{-3}$.
However, annihilations are particularly important at very high redshift ($z\gtrsim{}100$), where the particle density is very high. For this reason, even if the ionization fraction due to annihilations remains always very low, the Thomson optical depth is significantly high ($\tau{}_e\sim{}0.05$), even more than for LDM.

These results must be considered very optimistic upper limits. In fact we are assuming that nearly all the energy of the DM particle is immediately deposited into ionization or heating; whereas we expect that the electrons produced by neutralino annihilations Compton-scatter the CMB photons up to a energy $\sim{}1-10\,{}(1+z)$ MeV, which cannot be significantly absorbed by the intergalactic medium within a Hubble time (Chen \& Kamionkowski 2004).

\begin{figure}
\center{{
\epsfig{figure=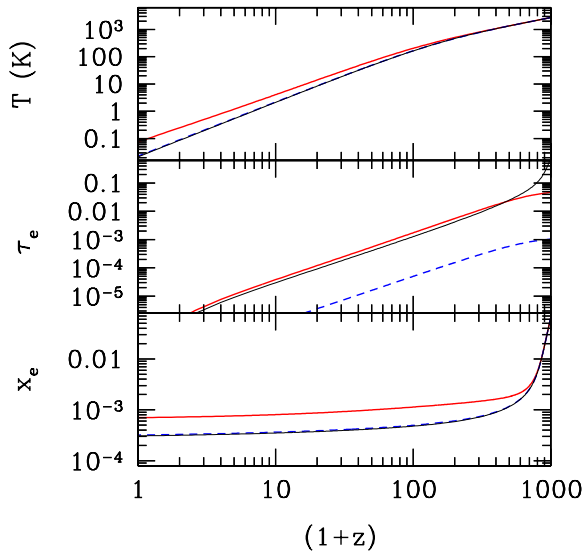,height=8cm}
}}
\caption{\label{fig:fig3} 
Ionized fraction (bottom panel), Thomson optical depth (central panel) and matter temperature (upper panel) as a function of redshift due to neutralinos for $\langle{}\sigma{}\,{}v\rangle{}=2\times{}10^{-26}$ (thick dashed line) and $10^{-24}$ cm$^3$ s$^{-1}$ (solid). In both the cases the neutralino mass is 100 GeV. The thin solid line  is the same as in Fig. 1.
} 
\end{figure}
\section{Warm dark matter: sterile neutrinos}
Sterile neutrinos are one of the most popular WDM candidates (Colombi, Dodelson \& Widrow 1996; Sommer-Larsen \& Dolgov 2001), even if Seljak et al. (2006) seem to exclude that they are the only component of DM on the basis of Ly-$\alpha{}$ forest power spectrum measurements. They can exist only if neutrinos have non-zero mass and mixing angles, as predicted by the standard oscillation theory (Dolgov \& Hansen 2002; see Dolgov 2002 for a complete review of sterile neutrino properties). There are many possible decay channels of sterile neutrinos (Dolgov 2002). In this paper we are interested on the radiative decay, i.e. the decay of a sterile neutrino into a lighter neutral fermion (such as an active neutrino) and a photon, because of its effects on the cosmic ionization and heating. 
From the comparison between the predicted background flux due to radiatively decaying sterile neutrinos and the hard X-ray background (Bauer et al. 2005), Mapelli \& Ferrara (2005) have established an upper limit of 14 keV for the sterile neutrino mass. This limit can now be lowered to $m_{\nu\,{}s}<$ 11 keV, adopting the relation between the mixing angle and the mass recently derived by Abazajian (2006). A stronger upper limit, $m_{\nu\,{}s}<$ 8.2 keV, has been derived from X-ray observations of the Virgo cluster (Abazajian 2006). Furthermore, Viel et al. (2005) derived a lower limit $m_{\nu\,{}s}>$ 2 keV from the study of matter power spectrum fluctuations. Then, sterile neutrino masses are allowed from 2 to 8 keV, a very narrow range.

The lifetime for sterile neutrino radiative decay is (Mapelli \& Ferrara 2005):
\q\label{eq:eq10}
\tau{}=\frac{512\,{}\pi{}^4}{9\,{}\alpha{}_{em}}\,{}G_F^{-2}m_{\nu\,{}s}^{-5}\sin{}^{-2}\theta{},
\nq
where $\alpha{}_{em}$ is the fine structure constant, $G_F$ the Fermi constant, $m_{\nu\,{}s}$ the sterile neutrino mass and $\sin{}\theta{}$ the mixing angle. To derive  $\sin{}\theta{}$ we adopt the following relation (Abazajian 2006):
\qa\label{eq:eq11}
\textrm{sin}^2\theta{}=2.5\times{}10^{-9}\left[\left(\frac{3.4\,{}\textrm{keV}}{m_{\nu\,{}s}}\right)\,{}\left(\frac{\Omega_{DM}}{0.26}\right)^{1/2}\right]^{1.626}\nonumber\\
\times{}\left\{0.527\,{}\textrm{erfc}\left[-1.15\,{}\left(\frac{T_{QCD}}{170\textrm{ MeV}}\right)^{2.15}\right]\right\}^{1.626},
\nqa
where $\Omega_{DM}$ is the dark matter density and $T_{QCD}$ the temperature of quark-hadron transition.

Assuming that all the DM is composed by sterile neutrinos and substituting equation (\ref{eq:eq10}) into equation (\ref{eq:eq1}) we derive through RECFAST the ionization and heating history  also for WDM particles (Fig. 4). Also sterile neutrinos start to play a role into the reionization and heating at redshift $z\sim{}100$, and their behavior is close (even if the global contribution is slightly lower) to that of LDM.
\begin{figure}
\center{{
\epsfig{figure=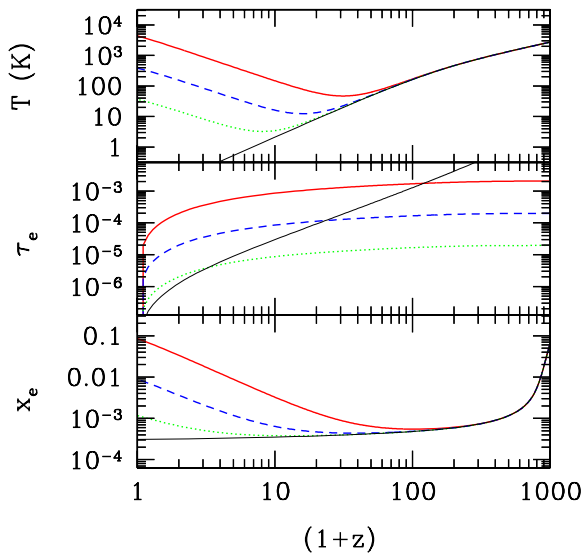,height=8cm}
}}
\caption{\label{fig:fig4} 
Ionized fraction (bottom panel), Thomson optical depth (central panel) and matter temperature (upper panel) as a function of redshift due to radiatively decaying sterile neutrinos of masses 2 (thick dotted line), 4 (dashed) and 8 keV (solid). The thin solid line  is the same as in Fig. 1.
} 
\end{figure}

\section{Effects on the CMB spectrum}
\begin{figure}
\center{{
\epsfig{figure=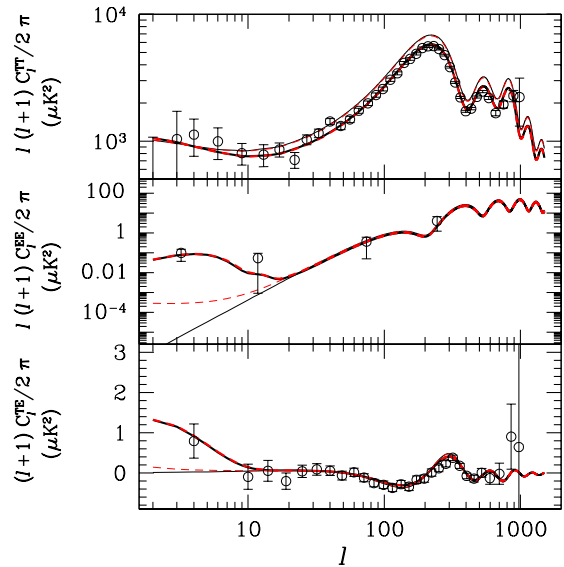,height=8cm}
}}
\caption{\label{fig:fig5}  
Temperature-temperature (top panel), polarization-polarization (central panel) and temperature-polarization (bottom panel) spectra. Thick lines indicate the CMB spectrum derived assuming Thomson optical depth $\tau{}_e=0.09$ and a sudden reionization model (consistent with the three year WMAP data); thin lines indicate the CMB spectrum derived assuming $\tau{}_e=0$. Dashed (solid) lines indicate the CMB spectrum obtained (without) taking into account the decays of 10 MeV LDM particles. The two thick lines, solid and dashed, appear superimposed, because the contribution of decaying particles (the dashed line) is completely hidden by the stronger effect of a sudden reionization with $\tau_e=0.09$. Open circles in all the panels indicate the three year WMAP data (Hinshaw et al. 2006; Page et al. 2006; Spergel et al. 2006).} 
\end{figure}
In the previous section we have shown that decaying DM, and especially LDM and sterile neutrinos, can modify the ionization fraction, with respect to the value due to relic electrons, already at high redshift. This fact should leave some imprint on the CMB spectrum (Chen \& Kamionkowski 2004; Pierpaoli 2004; Padmanabhan \& Finkbeiner 2005). To check whether these effects are measurable, we simulated the expected CMB spectrum in the case we take into account DM decays. This has been done by implementing our modified version of RECFAST in the version 4.5.1 of the public code CMBFAST (Seljak \& Zaldarriaga 1996; Seljak et al. 2003).

Fig.~5~shows~the~temperature~-~temperature~(TT), temperature~-~polarization~(TE) and polarization~-~polarization~(EE) spectra, in the case of 10 MeV decaying LDM (i.e. the particle for which we achieved the maximum contribution to the reionization among the considered ones), compared with the recent three year WMAP data (Spergel et al. 2006; Page et al. 2006). The contribution due to DM decays alone  is negligible. There is a sensible difference only in the lowest multipoles ($l<10$) of the EE spectrum. This effect can be seen in the central panel of Fig.~5, where the thin lines show the expected EE spectra by considering (dashed line) and neglecting (solid line) DM decays, respectively. This effect, small and concentrated at low multipoles, is justified by the fact that DM decays produce a very small Thomson optical depth ($\tau_e\lesssim{}0.01$) and that they are important especially at very low redshift, due to their long lifetime. Is such a modification of the EE spectrum measurable? If there are other sources of reionization besides DM decays (as it seems to be likely, considering the Thomson optical depth, $\tau_e=0.09^{+0.03}_{-0.03}$, measured by WMAP; Spergel et al. 2006), the influence exerted on the EE spectrum by the decaying DM would be completely hidden by the stronger effects due to these other reionizing sources. This can be seen in Fig. 5, where the thick lines show the TT/EE and TE spectra assuming  $\tau_e=0.09$ in the case with (dashed line) and without (solid line) DM decays. We found that the effects of DM decays are washed out by those of other reionizing sources also for lower values of $\tau_e$ consistent with the three year WMAP results (down to $\tau_e=0.06$, corresponding to a sudden reionization at $z\sim{}6$).

Because 10 MeV LDM particles produce the highest ionization fraction among the considered models, the effects on the CMB spectra due to other species of DM particles will be far more negligible. However, stronger effects on the CMB spectra can be due to annihilations of LDM particles (Zhang et al. 2006), which are not considered by this paper.

\section{Conclusions}
We examined the contribution to cosmic reionization and heating of different models of decaying/annihilating DM. In the case of quite light particles ($\lesssim{}10$ MeV) this contribution is significant.  Light particles (LDM, 
or sterile neutrinos)
become important at $z\sim{}100$, as they provide an
early  source of (partial) ionization
for the intergalactic medium.
They are expected to produce a Thomson optical depth $\tau{}_e\lesssim{}0.01$ ($\tau{}_e\lesssim{}0.001$) in the case of LDM particles (sterile neutrinos), which is smaller than the value derived from the three year WMAP data ($\tau_e=0.09$), but non-negligible. 
Changes in the matter temperature are also important, and the role of DM decays on the history of 21 cm emission should be investigated. On the contrary, heavier particles (gravitinos and neutralinos) do not have significant influence on reionization and heating.

This result could be crucial in distinguishing between light and heavy dark matter models, if new measures will be available of the reionization history and/or the behavior of the matter temperature (e.g. mapping the 21 cm emission at $z\sim{}10-50$). However, it is quite impossible to distinguish among different species of light DM particles, such as sterile neutrinos or LDM.

Finally, in the case of light particles, an early ($z\gg{}20$) increasing of the ionization fraction and of the baryon temperature could catalyse the production of H$_2$ and HD molecules, affecting the entire history of structure formation (Shchekinov \& Vasiliev 2004; Biermann \& Kusenko 2006).
On the contrary, no constraints on DM particles can be derived from their effects on the CMB spectra.

\section*{Acknowledgements}
We thank P.~Ullio, A.~Provenza, E.~Ripamonti and K.~Abazajian for useful discussions. We also acknowledge the Referee for his critical reading of the manuscript. E.P. is an ADVANCE fellow (NSF grant AST-0340648), also supported by NASA
grant NAG5-11489.


\onecolumn
\appendix


\begin{thebibliography}{}
\bibitem[\protect\citename{} ]{}Abazajian K., 2006, PhRvD, 73f, 3506
\bibitem[\protect\citename{} ]{}Ascasibar Y., Jean P., Boehm C., Kn\"odlseder J., 2006, MNRAS, 368, 1695, astro-ph/0507142
\bibitem[\protect\citename{} ]{}Bauer F. E., Alexander D. M., Brandt W. N., Schneider~D.~P., Treister~E., Hornschemeier~A.~E., Garmire~G.~P., 2004, AJ, 128, 2048 [B04]
\bibitem[\protect\citename{} ]{}Beacom J. F., Bell N. F., Bertone G., 2005, PhRvL, 94q, 1301
\bibitem[\protect\citename{} ]{}Beacom J. F., Y\"uksel H., 2006, submitted, astro-ph/0512411
\bibitem[\protect\citename{} ]{}Bertone G., Hooper D., Silk J., 2005, Phys.Rept., 405, 279
\bibitem[\protect\citename{} ]{}Biermann P. L., Kusenko A., 2006, PhRvL, 96L, 1301
\bibitem[\protect\citename{} ]{}Bode P., Ostriker J. P., Turok, N., 2001, ApJ, 556, 93
\bibitem[\protect\citename{} ]{}Boehm C., Hooper D., Silk J., Casse M., Paul J., 2004, PhRvL, 92j, 1301
\bibitem[\protect\citename{} ]{}Cass\'e M., Fayet P., 2005,  to appear in Proc. 21st IAP Colloquium "Mass Profiles and Shapes of Cosmological Structures", Paris 4-9 July 2005 (EAS Publications Series, G. Mamon, F. Combes, C. Deffayet, B. Fort eds.), astro-ph/0510490
\bibitem[\protect\citename{} ]{}Chen X., Kamionkowski M., 2004, PhRvD, 70d, 3502
\bibitem[\protect\citename{} ]{}Colafrancesco S., Profumo S., Ullio P., 2005, submitted to A\&{}A, astro-ph/0507575
\bibitem[\protect\citename{} ]{}Colombi S., Dodelson S., Widrow L. M., 1996, ApJ, 458, 1
\bibitem[\protect\citename{} ]{}Dermer C. D., Murphy R. J., 2001, in Exploring the gamma-ray universe. Proceedings of the Fourth INTEGRAL Workshop, 4-8 September 2000, Alicante, Spain. Editor: B. Battrick, 115
\bibitem[\protect\citename{} ]{}Dolgov A. D., 2002, Physics Reports, Volume 370, Issue 4-5, p. 333-535 (hep-ph/0202122)
\bibitem[\protect\citename{} ]{}Dolgov A. D., Hansen S. H., 2002, APh, 16, 339
\bibitem[\protect\citename{} ]{}Hansen S. H., Haiman Z., 2004, ApJ, 600, 26
\bibitem[\protect\citename{} ]{}Hinshaw G. et al. 2006, submitted to ApJ
\bibitem[\protect\citename{} ]{}Hooper D., Wang L.-T., 2004, PhRvD, 70f, 3506
\bibitem[\protect\citename{} ]{}Kasuya S., Kawasaki M.,  2004, PhRvD, 70j, 3519
\bibitem[\protect\citename{} ]{}Kasuya S., Kawasaki M., 2006, PhRvD, 73f, 3007
\bibitem[\protect\citename{} ]{}Kasuya S., Kawasaki M., Sugiyama N., 2004, PhRvD, 69b, 3512
\bibitem[\protect\citename{} ]{}Kasuya S., Takahashi F., 2005, PhRvD, 72h, 5015
\bibitem[\protect\citename{} ]{}Kawasaki M., Yanagida T., 2005, Physics Letters B, Volume 624, Issue 3-4, p. 162
\bibitem[\protect\citename{} ]{}Kn\"odlseder J. et al., 2005, A\&{}A, 441, 513
\bibitem[\protect\citename{} ]{}Mapelli M., Ferrara A., 2005, MNRAS, 364, 2
\bibitem[\protect\citename{} ]{}Nowakowski M., Rindani S. D., 1995, Phys.Lett., B348, 115
\bibitem[\protect\citename{} ]{}Ostriker J. P., Steinhardt, P., 2003, Sci, 300, 1909
\bibitem[\protect\citename{} ]{}Padmanabhan N., Finkbeiner D. P.,2005, PhRvD, 72b, 3508
\bibitem[\protect\citename{} ]{}Padmanabhan T., {\it Structure Formation in the Universe}, Cambridge University Press, 1995
\bibitem[\protect\citename{} ]{}Page L. et al. 2006, submitted to ApJ
\bibitem[\protect\citename{} ]{}Palla F., Salpeter E.E., Stahler S.W., 1983, ApJ 271, 632
\bibitem[\protect\citename{} ]{}Pierpaoli E., 2004,  Phys. Rev. Lett., 92, 031301
\bibitem[\protect\citename{} ]{}Sciama D. W., 1982, MNRAS, 198, 1
\bibitem[\protect\citename{} ]{}Seager S., Sasselov D. D., Scott D., 1999, ApJ, 523L, 1
\bibitem[\protect\citename{} ]{}Seager S., Sasselov D. D., Scott D., 2000, ApJS, 128, 407
\bibitem[\protect\citename{} ]{}Seljak U., Makarov A., McDonald P., Trac H., 2006, submitted to PRL, astro-ph/0602430
\bibitem[\protect\citename{} ]{}Seljak U., Sugiyama N., White M., Zaldarriaga M., 2003, PhRvD, 68h, 3507
\bibitem[\protect\citename{} ]{}Seljak U., Zaldarriaga M., 1996, ApJ, 469, 437
\bibitem[\protect\citename{} ]{}Shchekinov Y.A., Vasiliev E.O., 2004, A\&{}A 419, 19
\bibitem[\protect\citename{} ]{}Sommer-Larsen J., Dolgov A., 2001, ApJ, 551, 608
\bibitem[\protect\citename{} ]{}Spergel, D. N. et al., 2003, ApJS, 148, 175
\bibitem[\protect\citename{} ]{}Spergel, D. N. et al., 2006, submitted to ApJ
\bibitem[\protect\citename{} ]{}Viel M., Lesgourgues J., Haehnelt M. G., Matarrese S., Riotto A., 2005, PhRvD, 71f, 3534
\bibitem[\protect\citename{} ]{}Zhang L., Chen X., Lei Y.-A., Si Z., 2006, submitted to PRD, astro-ph/0603425
\end{thebibliography}
\end{document}